\documentclass{iau}
\usepackage{graphicx}

\title[IAUS337.ALFABURST] 
{Initial Results from the ALFABURST Survey}

\author[Surnis et al.]   
{M. P. Surnis$^1$, G. Foster$^{2,3}$, G. Golpayegani$^1$, A. Karastergiou$^2$, D. Lorimer$^1$, J. Chennamangalam$^2$, K. Rajwade$^1$, M. McLaughlin$^1$,
D. Agarwal$^1$, W. Armour$^2$, D. Werthimer$^3$, J. Cobb$^3$, A. Siemion$^3$, D. MacMahon$^3$, D. Gorthi$^3$ \and Pei Xin$^4$}

\affiliation{$^1$ West Virginia University, Morgantown, WV, USA.
\\ email: {\tt mayuresh.surnis@mail.wvu.edu} \\[\affilskip]
$^2$ University of Oxford, Oxford, UK.
$^3$ University of California, Berkeley, CA, USA.
$^4$ Xinjiang Astronomical Observatory, Urumqi, Xinjiang, China.}

\pubyear{2017}
\volume{337}  
\jname{Pulsar Astrophysics -­ The Next 50 Years}
\editors{P. Weltevrede, B.B.P. Perera, L. Levin Preston \& S. Sanidas, eds.}

\begin{document}

\maketitle

\begin{abstract}
Here, we present initial results from the ALFABURST radio transient survey, which is currently running in a commensal mode with the ALFA receiver at the Arecibo telescope. We observed for a total of 1400 hours and have detected single pulses from known pulsars but did not detect any FRBs. The non-detection of FRBs is consistent with the current FRB sky rates.

\keywords{surveys, (stars:) pulsars: general, methods: data analysis}
\end{abstract}


\section{Introduction}
Fast Radio Bursts (FRBs) are millisecond duration, intense bursts of radio emission originating at cosmological distances. To date, 23 FRBs have been detected [for an up-to-date list, see \cite[Petroff \etal\ (2016)]{Pet16}] with one [FRB 121102; \cite[Spitler \etal\ (2014)]{Spi14}], showing multiple bursts. The singular occurrence of most FRBs makes real-time detection and follow-up essential in order to achieve accurate localizations and to identify multi-wavelength counterparts.

ALFABURST is a commensal backend on the Arecibo L-band Feed Array (ALFA). It carries out real-time transient detection using a dedicated high performance computing configuration with field programmable gate arrays and graphics processing units. It processes 56 MHz bandwidth from each of the seven beams of the ALFA band with a sampling time of 256 $\mu$s.

\section{Transient Detection Pipeline}
ALFABURST is a component of SETIBURST (\cite[Chennamangalam \etal\ 2017]{Chen17}), a robotic, commensal, real-time backend deployed at the Arecibo telescope. It obtains the spectra from each beam and subjects them to radio frequency interference clipping, followed by de-dispersion of 8.4 s of data buffers over a trial dispersion measure (DM) range of 0 $-$ 10000 pc cm$^{-3}$. The de-dispersed time series for each trial DM are then smoothed to factors of 2 $-$ 16 in powers of 2. The smoothed time series are then searched for individual events crossing the threshold of 10 times the noise rms. Candidate FRBs are then passed to a coincidence filter, eliminating events seen in more than 3 beams simultaneously and a list of candidates along with the filterbank data for each candidate are saved to a local disk for further scrutiny. We are currently working on developing an artificial intelligence algorithm to categorize candidates automatically.  

\section{Results and Discussion}
\begin{figure}

\includegraphics[scale=0.39]{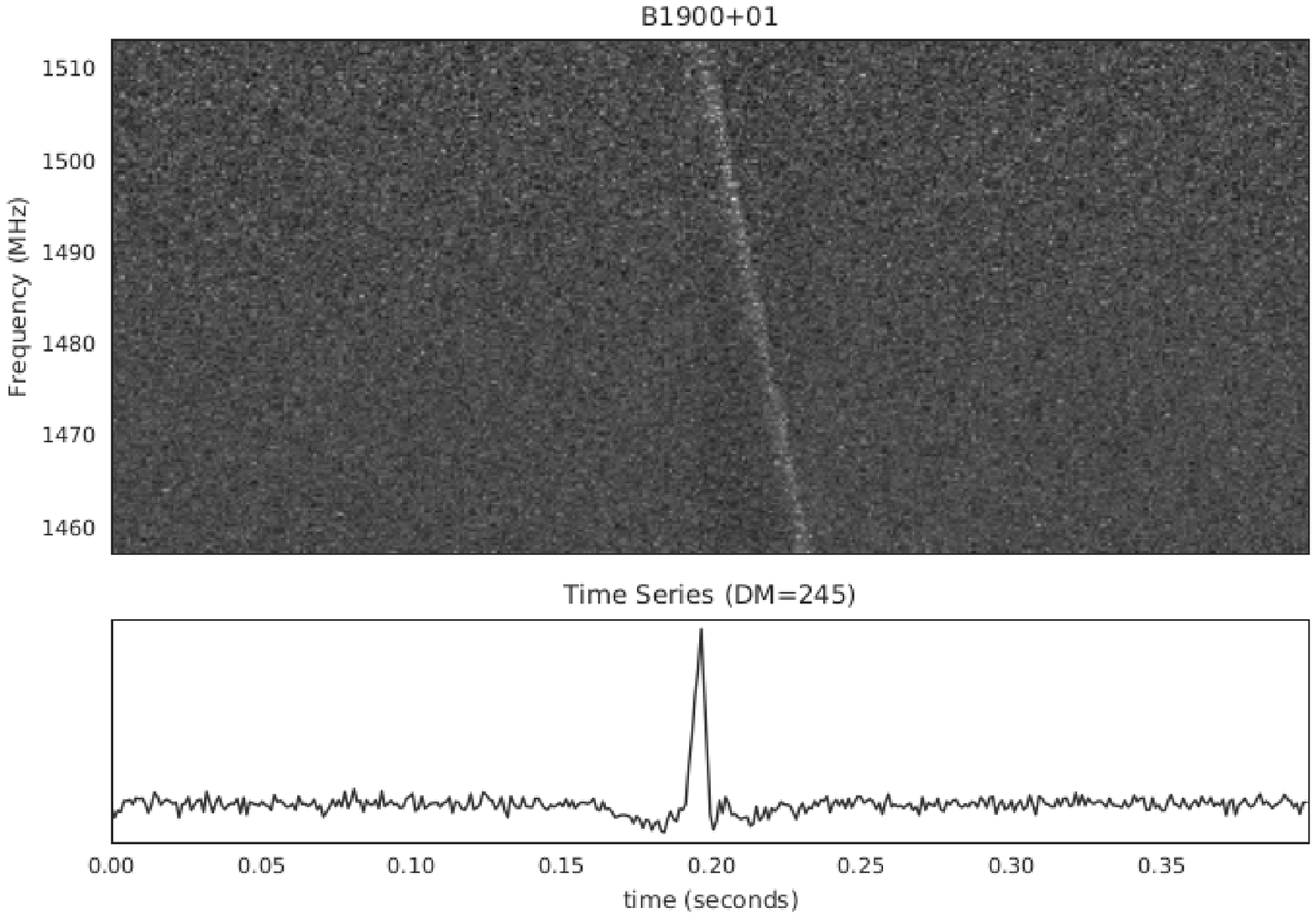}
\includegraphics[scale=0.095]{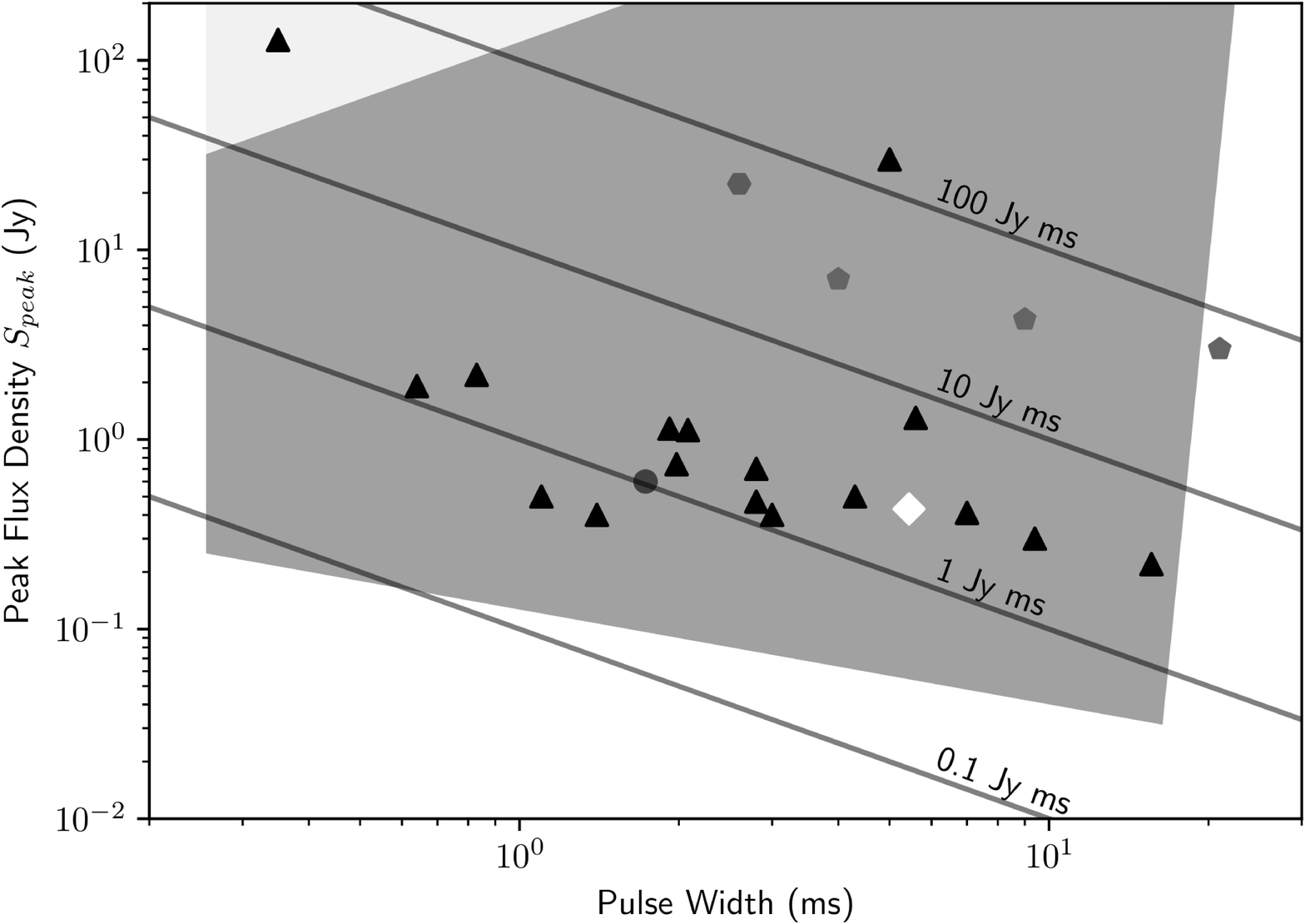}

\caption{\textit{Left:} Single pulse detected from PSR B1900+01. The grayscale plot shows the time-frequency structure while the bottom panel shows the pulse in the time series. \textit{Right:} Survey sensitivity (dark gray) as a function of pulse width. Automated RFI excision excludes narrow in width, bright FRBs such as FRB150807 (light gray). Previously detected FRBs from Parkes (triangles), GBT (circle), Arecibo (diamond), UTMOST (pentagons), and ASKAP (hexagon) are plotted for reference.}
\label{fig1}
\end{figure}

We did not detect any FRBs in the data taken through June 2017, with about 1400 hours of observing time. We have detected single pulses from known pulsars (see left panel in Figure \ref{fig1} for an example) in the observed region. We estimate the nominal sensitivity of the survey to be about 70 mJy for a pulse width of 10 ms. Figure \ref{fig1} (right panel) shows the sensitivity as a function of pulse width along with the 23 known FRBs for a comparison. Assuming the standard candle model (\cite[Lorimer \etal\ 2013]{Lor13}) and a volumetric FRB rate of 10$^{-3}$ per galaxy per year (\cite[Thornton \etal\ 2013]{Tho13}), we estimate a detection of at most 2 FRBs for the duration of observations. Given the uncertainties in the physical characteristics of known FRBs, our non-detection is consistent with the current models of the origin of the FRBs.

\end{document}